\documentclass[%
 reprint,
nofootinbib,
 amsmath,amssymb,
 aps,prl,
floatfix,
]{revtex4-2}

\allowdisplaybreaks
\newcommand{\Mp}{M_{\rm{P}}}

\usepackage{CJKutf8}
\usepackage[utf8]{inputenc} 
\usepackage{graphicx}
\usepackage{dcolumn}
\usepackage{bm}
\usepackage{xcolor}

\begin{document}

\preprint{APS/123-QED}

\title{Running primordial perturbations: Inflationary Dynamics and Observational Constraints}

\author{Richard Easther}
 \email{r.easther@auckland.ac.nz}
\affiliation{
 Department of Physics, University of Auckland, Private Bag 92019, Auckland, New Zealand}%

\author{Benedict Bahr-Kalus}
  \email{benedictbahrkalus@kasi.re.kr}
\author{David Parkinson}%
 \email{davidparkinson@kasi.re.kr}
 \affiliation{Korea Astronomy and Space Science Institute, Yuseong-gu, Daedeok-daero 776, Daejeon 34055, Korea}

\date{\today}

\begin{abstract}
Inflationary cosmology proposes that the early Universe undergoes accelerated expansion, driven, in simple scenarios, by a single scalar field, or inflaton. The form of the inflaton potential determines the initial spectra of density perturbations and gravitational waves.  We show that constraints on the duration of inflation together with the  BICEP3/Keck bounds on the gravitational wave background imply that higher derivatives of the potential are nontrivial with a confidence of 99\%. Such terms contribute to the scale-dependence, or running, of the density perturbation spectrum. We clarify the ``universality classes'' of inflation in this limit showing that a very small gravitational wave background can be correlated with a larger running. If pending experiments do not observe a gravitational wave background the running will be at the threshold of detectability if inflation is well-described at third-order in the slow roll expansion. 

\end{abstract}

\maketitle




Now forty years old, inflation \cite{Guth:1980zm} is the {\em de facto} description of the very early Universe. The clear consequences of generic inflationary models are well-verified: the Universe is spatially flat, almost homogeneous and isotropic and Gaussian, adiabatic perturbations \cite{PhysRevLett.49.1110,HAWKING1982295}  induce large scale correlations in the polarization and temperature of the microwave background \cite{starobinskii1983perturbation}. The one ambiguous observable is the primordial gravitational wave background. Constraints have steadily tightened~\cite{WMAP:2003elm,Planck:2018vyg,BICEP2:2018kqh,Tristram:2020wbi} and the  latest BICEP3/Keck data permits an amplitude of at most 4\% that of the density perturbations \cite{BICEP:2021xfz}.   A  gravitational wave background is often viewed as the ``smoking gun'' of inflation since known alternatives  do not generate  a detectable signal \cite{PhysRevD.69.127302,Ijjas:2015zma} but this is also true of many  inflationary models. Moreover,  algebraically simple slow-roll scenarios with large gravitational-wave signals must be ``protected'' by near-symmetries~\cite{Lyth:1996im}:  such models can be proposed  (e.g. \cite{Kachru:2003aw}) but nature need not employ them.

The amplitudes of the density and gravitational wave perturbations (expressed via their ratio, $r$), depend on the potential $V$ and its slope $V'$. The spectral index of the density perturbations $n_\mathrm{s}$ further involves the second derivative,  $V''$. Given a single field slow-roll prior, $n_\mathrm{s}$ and $r$ are inputs for the inflationary inverse problem: the {\em reconstruction} of the potential from observational data~\cite{Lidsey:1995np}.

We show that the latest BICEP3/Keck data implies that {\em all} viable implementations of slow-roll inflation with only $V$, $V'$ and $V''$ as free parameters produce more than 65 e-folds of inflation after astrophysically relevant perturbations leave the horizon, with 99\% confidence.  Without exotic post-inflationary physics, this is inconsistent with long-standing constraints  \cite{Dodelson:2003vq,Liddle:2003as,Adshead:2010mc,Munoz:2014eqa} so inflation can only terminate appropriately if higher derivatives are nontrivial or the potential is discontinuous. 

A nontrivial $V'''$ modifies the dynamics relative to that derived with only $V'$ and $V''$. For any  $n_\mathrm{s}$ and $r$ one can fix $V'''$ to yield a specified amount of inflation. However, this leads to  scale dependence in $V''$, contributing to the {\em running\/} of the spectral index,  $\alpha_{s} = d n_\mathrm{s} /d\ln{k}$ where $k$ is the comoving wavenumber. Experiments now under development are sensitive to  $r \gtrsim 10^{-4}$ \cite{CMB-S4:2016ple,Hazumi:2019lys}.  We show that if $r \lesssim 10^{-4}$,   it follows that $\alpha_{s} < -10^{-3}$, given three nontrivial slow-roll parameters. This is several times larger than $\alpha_{s}$ in simple models  \cite{Adshead:2010mc} and at the threshold of detection by upcoming  experiments. 

The key finding  is that all two-parameter single-field inflationary models are excluded with high confidence.  The analysis rests on the well-studied Hubble Slow Roll expansion \cite{Kinney:2002qn}. The full dynamical system  has apparent attractors \cite{Kinney:2002qn,Chongchitnan:2005pf} in the $\{n_s,r\}$ plane, issues with convergence and truncation (particularly in models with  discontinuities or an abrupt end to inflation, or many fields), and does not account for initial, transient, field velocities \cite{Vennin:2014xta}. However, none of these complexities impinge on our analysis. Present data limits us to the ``Low-$\epsilon$'' regime of the slow-roll hierarchy, excluding much of the attractor structure. Likewise generic scenarios beyond single-field slow-roll  necessarily have several free parameters, and thus cannot provide counterexamples that prevent the exclusion of  two parameter models. 

Given the measured value of $n_s$, we further show that tight constraints on $r$  imply a nontrivial running if the dynamics are treated at next-order in slow-roll. This formal linkage between the running and the duration of inflation is well known \cite{Adshead:2008vn,Malquarti:2003ia,Makarov:2005uh,Easther:2006tv,Adshead:2008vn,Adshead:2010mc}, and we clarify the understanding of inflationary universality classes in this limit \cite{Mukhanov:2013tua,Roest:2013fha,Creminelli:2014nqa,CMB-S4:2016ple}.  Correlated expectations for $r$ and $\alpha_\mathrm{s}$  depend on the  truncated slow-roll hierarchy but  three-parameter slow-roll is now the simplest feasible scenario. Excitingly, this linkage between $r$ and $\alpha_s$ presents a feasible target for future astrophysical measurements.

\vskip 12pt

\noindent {\em Two Parameter Slow-Roll Models:} 
Single-field inflationary scenarios are governed by the Einstein-Klein-Gordon equations,
\begin{align} 
&H^2 =\frac{1}{3 \Mp^2}\left( \frac{\dot\phi^2}{2} + V(\phi)\right) \, , \label{eq:hsqrd} \\  
&\dot{H} = -\frac{1}{2\Mp^2} \dot{\phi}^2\, , \label{eq:hdot} \\
&\ddot{\phi} + 3 H \dot \phi + V'(\phi) =0 \, ,
\end{align}
where the symbols have their usual meanings and we use the reduced Planck mass, $\Mp$. During the accelerated phase $\phi$ evolves monotonically and is thus a ``clock''. Eq.~{\ref{eq:hdot}} can be rearranged to show that $dH/d\phi$ is proportional to $-d\phi/dt$, so
\begin{equation} \label{VHSR}
   V(\phi) =  \frac{3 \Mp^2}{2} H(\phi)^2 -  \Mp^4 H'(\phi)^2.  
\end{equation}

For the purposes of parameter counting, we assume that the Potential and Hubble Slow Roll formulations are interchangeable. The Hubble Slow Roll hierarchy~\cite{Kinney:2002qn} provides a more succinct account of the dynamics,
\begin{eqnarray}\label{epsphi}
\epsilon(\phi) & \equiv &
2\Mp^{2}\left[\frac{H'(\phi)}{H(\phi)}\right]^{2},\\
^{\ell}\lambda_{H} & \equiv &
\left(2\Mp\right)^{\ell}\frac{(H')^{\ell-1}}{H^{\ell}}
\frac{d^{\ell+1}H}{d\phi^{(\ell+1)}};\;\ell \geq 1.
\label{eq:lgen}
\end{eqnarray}
with the convention that $\eta = {}^{1}\lambda_{H}$ and
$\xi = {}^{2}\lambda_{H}$. The number of e-folds that will elapse before inflation ends is $N=-\ln(a/a_{\mathrm{end}})$, where $a_{\mathrm{end}}$ is the scale factor as inflation completes. Noting $H=\dot{a}/a$, 
\begin{equation}
    \frac{dN}{d\phi} = \frac{1}{\Mp} \frac{1}{\sqrt{2\epsilon}} \, ,
\end{equation}
the ``flow equations'' are 
\begin{eqnarray} \label{flow1}
\frac{d\epsilon}{dN} & = & 2\epsilon(\eta-\epsilon),\\
\frac{d\eta}{dN} & = & -\epsilon\eta+ \xi\, , \\ \label{eq:flowellfull}
\frac{d\,{}^{\ell}\lambda_{H}}{dN} & = &
[(\ell-1)\eta-\ell\epsilon]\times {}^{\ell}\lambda_{H}+{}^{\ell+1}\lambda_{H} \, ,
\end{eqnarray}
where $N$ is now the independent variable. 
Accelerated expansion occurs when $\ddot{a}>0$ or equivalently $\epsilon<1$. If ${}^{\ell}\lambda_{H} =0 $ for all $\ell \ge M$ at some $\phi_0$ the system remains closed as it evolves \cite{Hoffman:2000ue,Kinney:2002qn,Liddle:2003py}, with $M$ nontrivial slow-roll parameters. The  amplitude of the potential is a further free parameter but scales out of the  dynamics.

\begin{figure}
    \centering
    \includegraphics[width = \columnwidth]{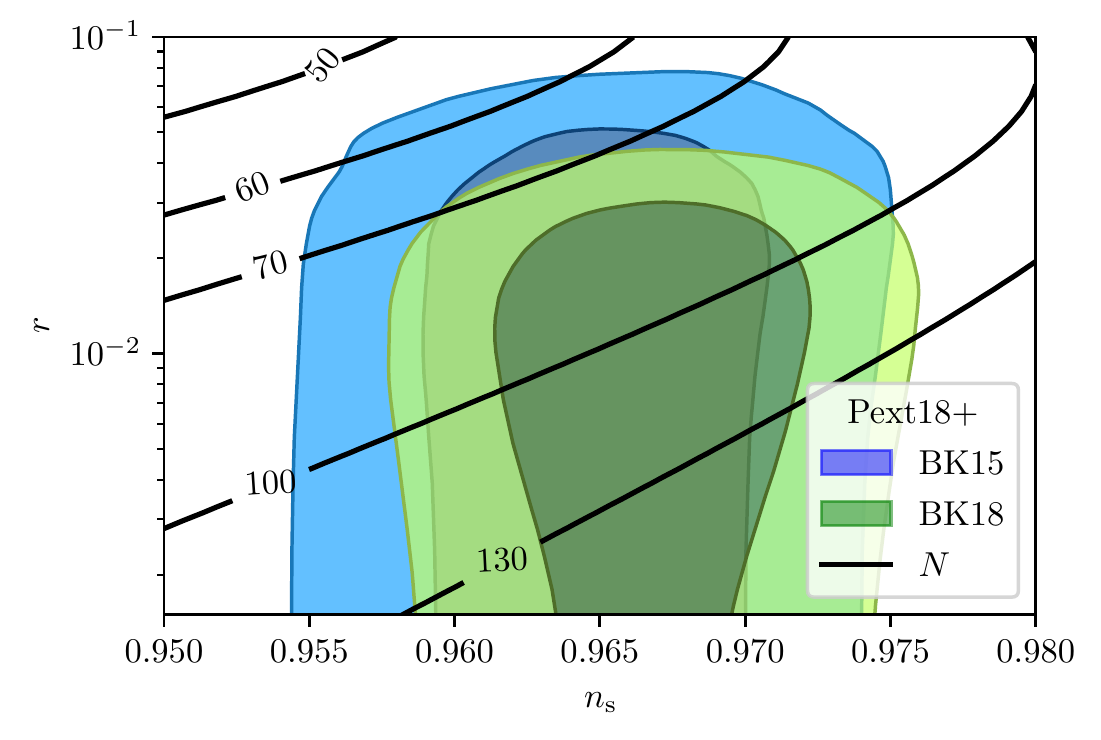}
   \caption{Likelihood contours in $n_\mathrm{s}$ and $\ln{r}$ from the BK15 (blue) and BK18 (green) datasets, in combination with  PlanckTTTEEE+lowE+lensing+BAO results~\cite{BICEP2:2018kqh,BICEP:2021xfz}. Shaded regions denote the 66\% and 95\% posteriors.
    Black contours indicate the number of e-folds $N$ that take place after the pivot leaves the horizon with a two-term slow-roll hierarchy.
    \label{fig:ns_r_N_cornerplot}}
\end{figure}

The foregoing treatment is exact but key observables are expressed in the slow-roll {\em approximation}, or 
\begin{eqnarray}\label{ns}
n_{s} & = & 1+2\eta  -
4\epsilon-2(1+\mathcal{C})\epsilon^2- \frac{1}{2}(3-\mathcal{C})\xi,
\\ \label{r}
r & = & 16\epsilon[1+2C(\epsilon-\eta)],\\
\alpha_{s} & = & -\frac{1}{1-\epsilon} \frac{d \phi}{dN} \frac{d n_\mathrm{s}}{d\phi}
\end{eqnarray}
where $C = -2+\ln 2+\gamma$, $\mathcal{C} = 4(\gamma+\ln 2)-5$, and $\gamma$ is the
Euler-Mascheroni constant. Finally, $dN/d\ln k = -1/(1-\epsilon)$ is the rate at which modes leave the horizon and $\epsilon \rightarrow 0$ in the de Sitter limit where $H$ is constant.  

\begin{figure}
    \centering
    \includegraphics[width = \columnwidth]{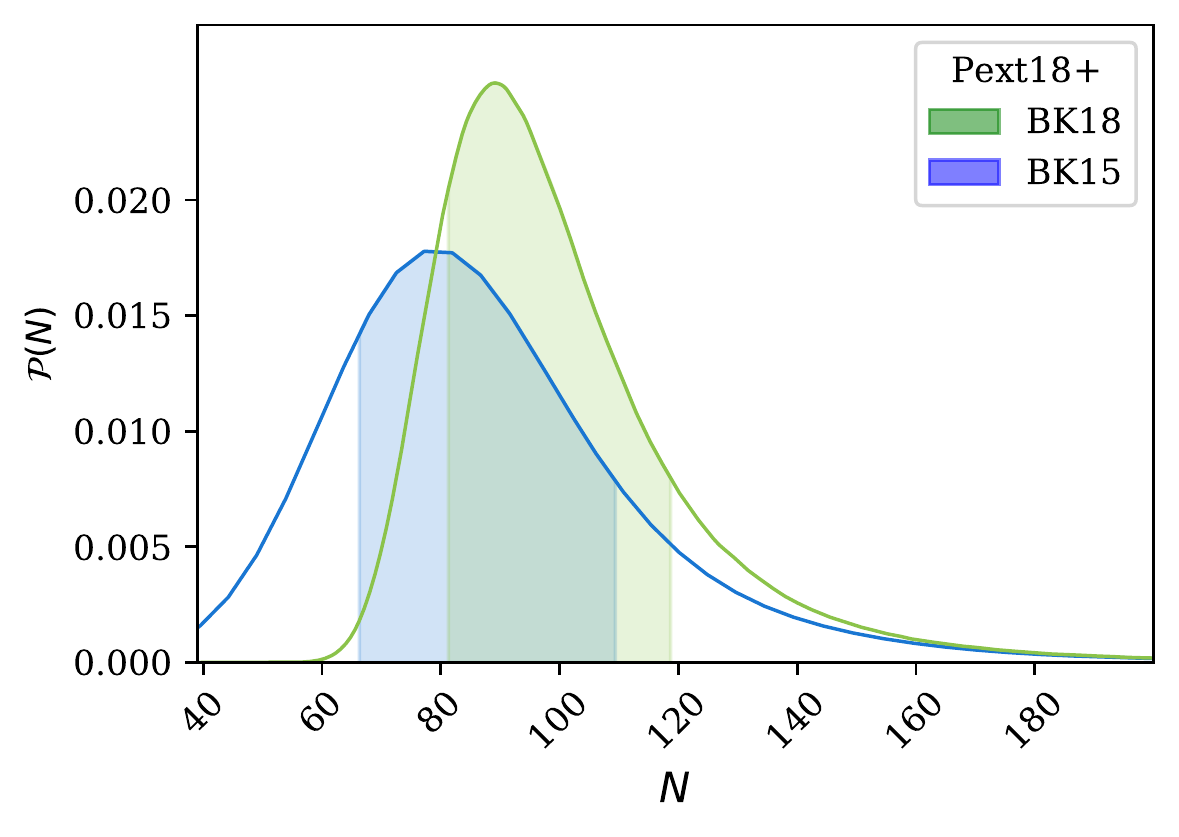}
    \caption{The posterior for $N$ with a two-term slow-roll hierarchy (as in Fig.~\ref{fig:ns_r_N_cornerplot}) with a uniform prior on $r$; for a logarithmic prior the distribution is roughly constant at larger $N$.  
} 
    \label{fig:BK18_Nfolds_marginalised_posterior}
\end{figure}

A two-term hierarchy maps $n_\mathrm{s}$ and $r$ to an inflationary trajectory. Fig.~\ref{fig:ns_r_N_cornerplot} shows the constraints on $n_\mathrm{s}$ and $\ln{r}$ derived from the BK15 \cite{BICEP2:2018kqh} and BK18~\cite{BICEP:2021xfz} datasets (published in 2018 and 2021, respectively), together with Planck and Baryon Acoustic Oscillation data  overlaid with the duration of inflation computed with two slow-roll terms.  Fig.~\ref{fig:BK18_Nfolds_marginalised_posterior} shows the  marginalised distributions for $N$; BK18 yields $P(N<65) \approx 0.0024$. Provided the post-inflationary universe is not dominated by matter whose stiffness exceeds that of radiation, $N<65$ is a generic bound on the amount of inflation after the pivot leaves the horizon \cite{Liddle:2003as}. Subject to this proviso on the equation of state,  all inflationary models described by the first two slow-roll parameters are now excluded.

This advance arises from tightening constraints on {\em both} $n_\mathrm{s}$ and $r$. A spectral index of less than 0.95 was consistent with the full WMAP datset \cite{WMAP:2012fli} and inflation ends ``on time'' for smaller $n_\mathrm{s}$ without additional curvature in the potential. Consequently,  better measurements of $n_\mathrm{s}$ combine with tighter bounds on the polarization to yield this result. Note too that this analysis implicitly assumes a ``ski-run'' inflationary potential with a smooth approach to regular expansion. Scenarios in which inflation abruptly terminates also require additional parameters, albeit outside the Hubble Slow Roll expansion.

\begin{figure}
    \centering
    \includegraphics[width = \columnwidth]{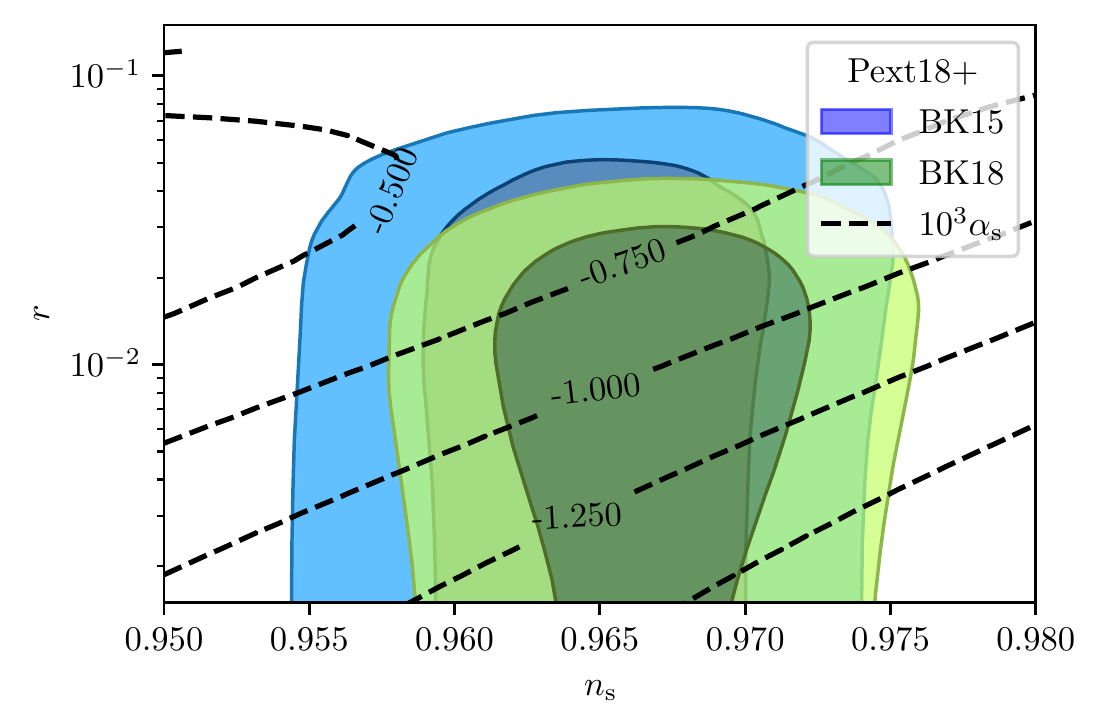}
    \caption{Contours of constant $\alpha_\mathrm{s}$ in the $n_\mathrm{s}-\ln{r}$ plane with  a three-term slow-roll hierarchy and $\xi$ set to give $N=55$ when the pivot leaves the horizon,  with the constraints from Fig.~\ref{fig:ns_r_N_cornerplot}.\label{fig:alpha}}
\end{figure}

\begin{figure*}
    \centering
    \includegraphics[width = \linewidth]{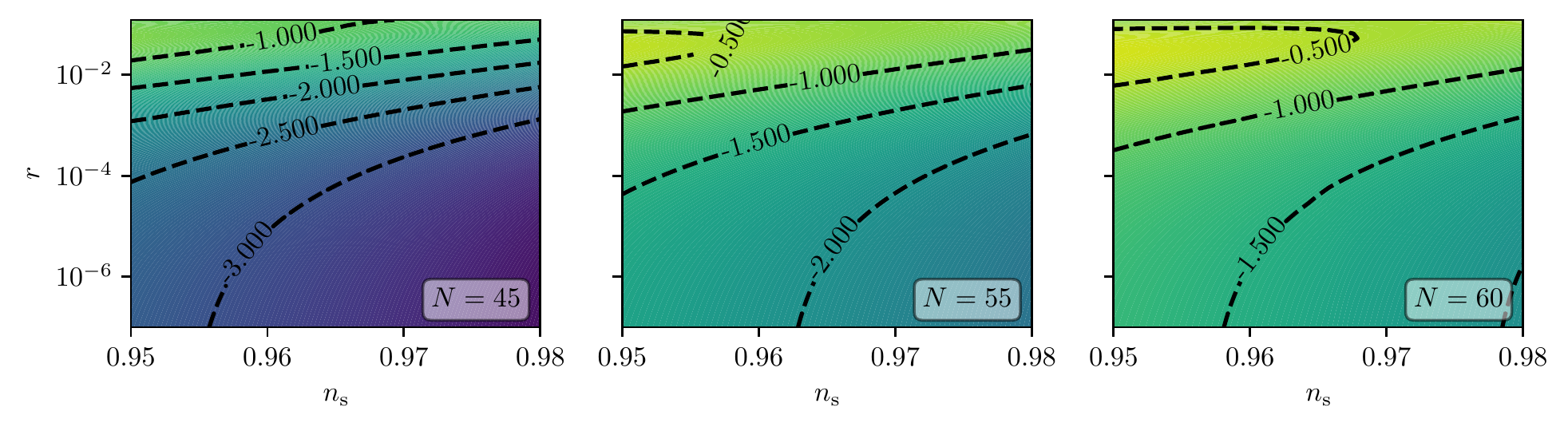}
           \caption{The running $\alpha_\mathrm{s}$ ($\times 10^3$) is plotted in the $n_\mathrm{s}-\ln{r}$ plane for $N = 45,55$ and $60$, assuming a three-parameter slow-roll hierarchy. When $r\lesssim 10^{-4}$ we see that $\alpha_\mathrm{s} <-10^{-3}$ for all values of $n_\mathrm{s}$ consistent with presently available data.  \label{fig:isoalpha}}
\end{figure*}

\vskip 12pt

\noindent {\em Running and the End of Inflation} We now extend the Hubble Slow Roll expansion to third order, so that  $\xi$ is non-zero.  This can increase the scale-dependence of $\eta$, as  $\alpha_\mathrm{s} \approx -2\xi$ when $\epsilon$ is small.  Fig.~\ref{fig:alpha} overplots the $n_\mathrm{s}$ and $r$ constraints with contours showing the running resulting from choosing $\xi$ such that $N=55$ when the pivot leaves the horizon. The running is  generically larger than in ``standard'' inflationary models \cite{Adshead:2010mc} but  still well  inside recent constraints; e.g. $d n_\mathrm{s}/d \ln k = -0.006 \pm 0.013$  \cite{Planck:2018vyg}. 

This  adds nuance to  statements that $n_\mathrm{s} -1 \sim -1/N$ and  $\alpha_\mathrm{s} \sim 1/N^2$ which hold empirically for many simple  models \cite{Adshead:2010mc}.  These expectations have been formalised in the Potential Slow Roll expansion \cite{Mukhanov:2013tua,Roest:2013fha,Creminelli:2014nqa,CMB-S4:2016ple}, leading to what are sometimes referred to as ``universality classes'' \cite{Roest:2013fha}. In this framework  $\epsilon_V = M_P^2 (V'/V)^2/2$, $\eta_V = M_p^2 V''/V$ and $\xi_V = M_p^4 V'V'''/V^2$   and
\begin{equation}
\frac{d \epsilon_V}{dN} \approx {M_P^4} \left(\frac{V'}{V}\right)^2 \left[ \frac{V''}{V} - \left(\frac{V'}{V}\right)^2\right] \, , 
\end{equation} 
We write $n_\mathrm{s}-1 = -a/N$, where $a$ is a constant a little larger than unity. Dropping higher order terms and accounting for the difference between $\eta$ and $\eta_V$ we can set this equal to Eq.~\ref{ns}, or $n_\mathrm{s} \approx 1-6\epsilon_V + 2\eta_V$ to find a differential equation for $\epsilon_V(N)$ (e.g. \cite{Creminelli:2014nqa}).  In the low $r$ limit the solution has the form $\epsilon_V \sim 1/(AN^a)$ where $A$ is a large constant. Physically,  this  ensures that $\eta_V$ and $\epsilon_V$ are tightly correlated even when $\epsilon_V \ll \eta_V$. However if $r \lesssim |n_\mathrm{s}-1|^2$ it would seem that  $\xi_V$  cannot be self-consistently ignored, since it contributes to the scale dependence of $\eta_V$ via
\begin{equation}
\frac{d \eta_V}{dN} \approx  {M_P^4}  \left[ \frac{V'}{V} \frac{V'''}{V} - \left(\frac{V'}{V}\right)^2 \frac{V''}{V} \right]  \, .
\end{equation}
and the second term can be far smaller than the first. 

This regime corresponds to the Low-$\epsilon$ limit of the Hubble Slow Roll hierarchy, and with three terms 
\begin{equation}
\frac{d \eta}{dN} \approx \xi \, , \qquad 
\frac{d \xi}{dN}   \approx \xi \eta \, .
\end{equation} 
These equations can be solved~\cite{Adshead:2008vn}, showing 
\begin{equation} \label{xiN}
    \xi(N) = \frac{\eta(N)^2 - \eta_\star^2}{2} + \xi_\star  
\end{equation}
where the star subscript denotes a value at the pivot.  To a good approximation   $\eta(N) = \eta_\star - \xi_\star \Delta N$ for astrophysically relevant modes,  where $\Delta N$ is the number of e-folds after the pivot leaves the horizon; the full  solution for $\eta(N)$ in this limit is the ``2-parameter, {Low-$\epsilon$}'' model of Ref.~\cite{Adshead:2008vn}.   In particular, the relationship $\xi \sim |n_\mathrm{s}-1|^2 \sim 1/N^2$ is supplemented by an additive constant in the Low-$\epsilon$ limit. Physically, this yields a near-inflexion point in the potential, where both $\epsilon$ and $n_\mathrm{s}-1$ are necessarily very small.

\begin{figure}
    \centering
    \includegraphics[width = 
    \columnwidth]{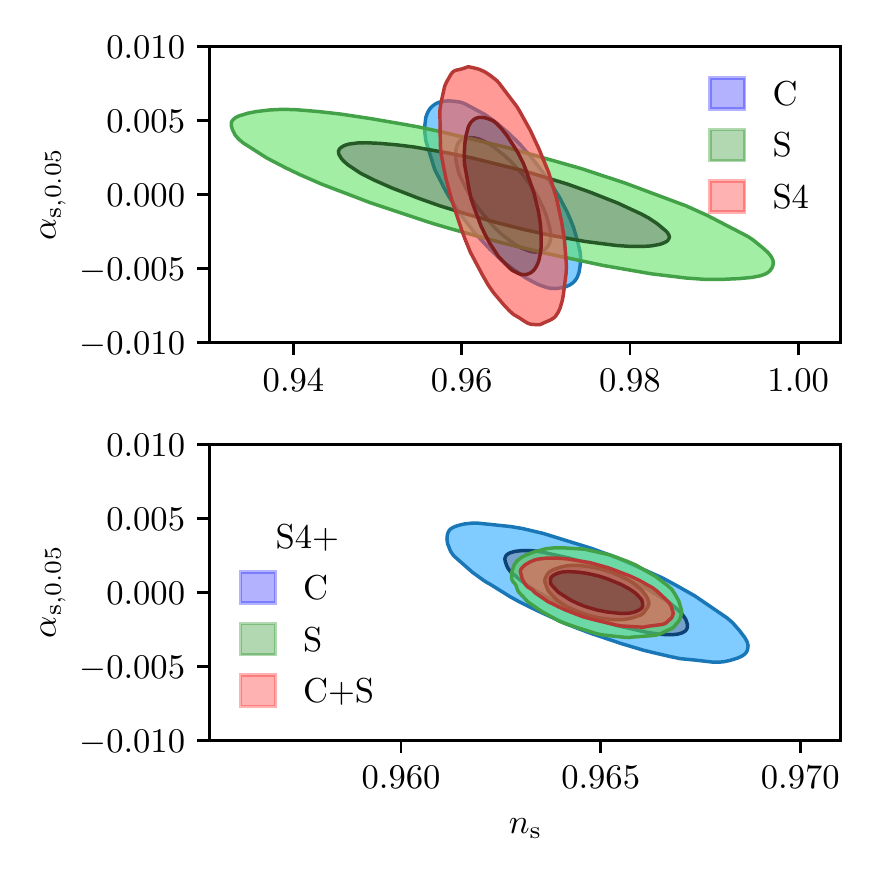}
    \caption{Forecast $n_\mathrm{s}$-$\alpha_\mathrm{s}$ constraints with CMB-S4, SPHEREx [S], CHIME [C]. The best combination promises to measure $\alpha_\mathrm{s}$ to about $2.2\times 10^{-3}$ at 95 \% confidence \cite{paper2}.}
    \label{fig:ns_alpha_forecasts}
\end{figure}

\vskip 12pt 

\noindent {\em Future Prospects:}   Recalling that $r\sim (V'/V)^2$, we can identify three regimes; $V' > V''$, $V' \sim V''$ and $V'\ll V''$ (with $M_p=1$). The first requires $r\gtrsim 0.01$ and is  close to being ruled out; the second is eliminated if {$r\lesssim 10^{-4}$}, a threshold which  will be within reach by 2030 \cite{CMB-S4:2016ple,Hazumi:2019lys}. 

If a primordial gravitational wave background is not detected in the coming decade, any viable single-field  model will satisfy $V' \ll V''$ and is thus squarely inside the Low-$\epsilon$ regime. Fig.~\ref{fig:isoalpha} shows the likely values of $\alpha_\mathrm{s}$ on the $n_\mathrm{s}-\ln{r}$ plane for three different choices of the total number of e-foldings.  If $r\lesssim 10^{-4}$ then $\alpha_\mathrm{s} <-10^{-3}$ for any self-consistent three-parameter scenario.  

Fig.~\ref{fig:ns_alpha_forecasts} shows the individual and combined limits on $|\alpha_\mathrm{s}|$ expected from CHIME \cite{chime} and  SPHEREx \cite{Dore:2014cca}, together with  CMB-S4 \cite{CMB-S4:2016ple}. Each experiment measures $\alpha_\mathrm{s}$ with an accuracy of, at best, $5\times 10^{-3}$ but their combined sensitivity is similar to the expected running if $r\lesssim 10^{-4}$. All these experiments aim to provide results by 2030. Consequently, if the early Universe passed through an accelerated phase the simplest currently viable inflationary models suggest that we can hope to have evidence that either $r$ or $\alpha_\mathrm{s}$ is non-zero in a decade from now.

\noindent \vskip 12pt

\noindent {\em Discussion} We have updated the priors on scalar field inflation using the latest data:  models specified by only  $V'$ and $V''$ at the pivot  do not lead to a self-consistent inflationary era, at a 99\% confidence level.  There is a clear correlation  between small $r$ and large $\alpha_s$  at third order in slow-roll. That said, it does not hold generically; even in slow-roll, if ${}^{3}\lambda_H$ is nontrivial  $\xi$ and  the running can be small at the pivot. Moreover there are further counterexamples which cannot be easily described within the Hubble Slow Roll hierarchy, e.g. multifield models, and those with discontinuous or modulated potentials.  

The current observational roadmap will investigate the range  $10^{-4}\lesssim r \lesssim 10^{-2}$ and $|\alpha_\mathrm{s}| \gtrsim 10^{-3}$ in the coming decade. Without a detection of the gravitational wave background there will be real pressure on the relationship between $\alpha_\mathrm{s}$, $r$ and $N$ highlighted here. Consequently, even a null result will significantly constrain what is now  the simplest  viable inflationary model in terms of parameter count and  qualitative complexity. 

This analysis also illuminates inflationary universality classes for very small  $r$, which arise from treating expressions for $n_\mathrm{s}$ as differential relationships. Conversely, the  Hubble Slow Roll parameters are akin to Taylor coefficients and  the  ``flow equations'' describe their running \cite{Kinney:2002qn}. The first two  terms set $\sqrt{r}$ and $|n_\mathrm{s}-1|$ but near an extremum of $V(\phi)$ (or $H(\phi)$, since $V'=0$ implies $H'=0$) $r \ll 1$ and $V'''$ is  the next-to-leading order term.

A scenario in which $V'$ is very small and $V'''$ is significant is most naturally an inflexion-point model. Interestingly, hilltop potentials of the form $V\sim V_0 - V_2 \phi^2 - V_4 \phi^4$ struggle to generate low values of $r$, given present constraints on $n_\mathrm{s}$  \cite{Martin:2013tda}.  In addition, for fixed $n_\mathrm{s}$ there is an inverse correlation between $\alpha_\mathrm{s}$ and $N$, depending on  the overall inflationary scale \cite{Liddle:2003as,Easther:2006tv} and the possibly complicated and nonlinear physics of the post-inflationary universe \cite{Shtanov:1994ce,Kofman:1997yn,Lozanov:2016hid,Kofman:1997yn,Jedamzik:2010dq,Easther:2010mr,Musoke:2019ima,Hasegawa:2019jsa}.  This overall discussion could be further sharpened by adopting  a Bayesian model comparison framework \cite{Norena:2012rs,Planck:2018vyg,Handley:2019fll}, and these considerations illuminate the viable forms of the inflationary potential. Note too that if the inflationary patch of the potential is small it is more likely that models will be sensitive to the initial spatial configuration of the inflaton  \cite{Goldwirth:1989pr,East:2015ggf,Clough:2016ymm}. 

In summary, all inflationary models fully described at second order in the Hubble slow roll expansion are now excluded by observational data with high confidence. This comes twenty years after the first nontrivial limits on inflationary models were delivered by WMAP  \cite{WMAP:2003elm,WMAP:2003syu}, and marks a  significant advance in the ability to constrain inflation. Moreover,  three-parameter slow-roll models, now the simplest scenarios (in terms of parameter count and qualitative complexity), exhibit a correlation between the gravitational wave amplitude and the running. This will be testable over the coming decade, and either a verification or a null result would represent major progress.

\section*{Acknowledgements}

\begin{acknowledgments}

RE  acknowledges support from the Marsden Fund of the Royal Society of New Zealand. BBK and  DP are supported by the project \begin{CJK}{UTF8}{mj}우주거대구조를 이용한 암흑우주 연구\end{CJK} (``Understanding Dark Universe Using Large Scale Structure of the Universe''), funded by the Ministry of Science of the Republic of Korea.  We analysed BICEP3/Keck chains \cite{BICEP2:2018kqh,BICEP:2021xfz} using \verb~ChainConsumer~\cite{Hinton2016} and plotted our results using \verb~Matplotlib~ \cite{Hunter:2007}. Plots in this paper were constructed by resampling chains made available by the BICEP/Keck collaboration.
\end{acknowledgments}

%



\end{document}